\documentstyle[art12]{article}
\begin{document}

\newcommand{\bean}{\begin{eqnarray*}}
\newcommand{\eean}{\end{eqnarray*}}
\newcommand{\ed}{\end{document}}
\newcommand{\pr}{\prime}
\newcommand{\ppr}{\prime\prime}
\newcommand{\cE}{{\tilde E}}
\newcommand{\vphi}{{\varphi}}
\newcommand{\oO}{O(k^{-1})}
\newcommand{\be}{\begin{equation}}
\newcommand{\ee}{\end{equation}}
\newcommand{\barr}{\begin{array}}
\newcommand{\earr}{\end{array}}
\newcommand{\bea}{\begin{eqnarray}}
\newcommand{\eea}{\end{eqnarray}}
\newcommand{\pa}{\partial}
\newcommand{\xx}{\hbox{}^*_*}
\newcommand{\sds}{\subset\hskip - 1em +}

\title{ Realization of the Dirac bracket algebras through first class
functions and quantization of constrained systems.}
\author{A.V.Bratchikov \\ Kuban
State Technological University,\\ 2 Moskovskaya Street, Krasnodar,
350072, Russia\\ E-mail:bratchikov@kubstu.ru} \date {March,\,2002}
\maketitle

\begin{abstract}
It is shown that a Dirac bracket algebra is isomorphic to
the original Poisson bracket algebra of
first class functions subject to first class constraints.
The isomorphic image of the Dirac bracket algebra in the star-product
commutator algebra is found.
\end{abstract}
\smallskip



{\bf1.}
Almost all the schemes of quantization of
dynamical systems with second class constraints are based on conversion
of original constraints into first class ones.
In the Dirac bracket approach  \cite {D} this conversion is
achieved by modification of Poisson brackets.
In the BRST method one introduces auxiliary
variables and constructs the BRST charge which is first class.Auxiliary
variables are also used in the conversion scheme of \cite {BFF}.

In this paper we present the approach where second class constraints
are converted into first class ones without using modification
of Poisson brackets or auxiliary variables.We
construct the algebra with respect to the original
Poisson bracket which is isomorphic
to the Dirac bracket algebra.
It is a quotient of the Poisson bracket algebra of first class
functions.
This algebra is singled out by covariant conditions and
includes only first class constraints.

A Dirac bracket algebra can be treated as the original Poisson bracket
algebra of the functions on constraint surface
\cite {S,L,FLS} (see also \cite {F} and
references therein ).However to apply this realization
one needs to know a solution to constraint equations.

The new realization enables us to  replace
quantization of the Dirac bracket algebras by
quantization of first class functions subject to first class
constraints.The latter seems to be more preferable.
An operator version of this approach was used for
quantization of two-dimensional coset
conformal field theories \cite {B}.

   It is known (see \cite {St} for a review) that the Poisson
bracket algebra is isomorphic to a quotient of the  corresponding
star-product commutator algebra.Using this fact and the connection
between the Dirac and Poisson bracket algebras we obtain realization of
the Dirac bracket algebra in terms of the star-product commutator
algebra.

We consider only systems with second
class constraints bearing in mind that a system with first
class constraints and gauge fixing is second class (see e.g. \cite
{GT}).

\bigskip

{\bf 2.}In this section we review the Dirac bracket approach
\cite {D} and introduce notations.
Let $M$ be a phase space with the phase variables $\eta_n,\,n=1...2N,$
and the Poisson bracket \bean \label{U} [\eta_m,\eta_{n}
]=\omega_{mn}(\eta). \eean Let $H(\eta)$ be the original hamiltonian and
$\varphi_j(\eta), j=1...2J,$ the second class constraints \bean
\label{U} det [\varphi_j,\varphi_{k} ]|_{\varphi=0} \ne 0.  \eean

The dynamic of the system under consideration is described by the
Hamilton equations
\bea
\label{h1}
\frac {d} {dt}
\eta_n = [\eta_n,H^T ],
\qquad
\varphi_j=0.
\eea
Here $H^T=H+{\lambda_j\varphi_j}.$
Functions $\lambda_j=\lambda_j(\eta)$ are defined by
the  equation \bea \label  {hh12}
[H^T, \varphi_j]|_{\varphi=0}=0.
\eea

Using (\ref {hh12}) one can write equations
(\ref{h1})
as
\bea \label{h3}
\frac {d} {dt}
\eta_n = [\eta_n,H^T ]_D,
\qquad \varphi_j=0.
\eea
Here the Dirac bracket was introduced
\bean \label{Dbr}
[g,f]_D=[g,f]- [g,\varphi_{j}]c_{jk}[\varphi_{k},f],\qquad
c_{jk}[\varphi_{k},\varphi_{l}]=\delta_{jl} .
\eean

Let $A=C^\infty(M)$ be the algebra of smooth functions on $M.$
One can associate with $A$ the Lie algebras $A_P$ and $A_D$
with the Poisson and Dirac brackets respectively.
Let $\Phi\subset
A$
be the subspace of the functions
which vanish on constraint surface
\bean \label{Did1}
\Phi =\{u\in A\,|\,u=u_j(\eta){\varphi_j}\}.
\eean
For $g\in A$
\bean \label{} [g,{\varphi_j}]_D=0.
\eean
From this it follows that for
$u\in \Phi,g\in A $
\bean \label{} [u,g]_D\in \Phi.
\eean
Hence $\Phi$ is an ideal of $A_D$
and the quotient $A_D/\Phi$ is an algebra.

It was observed by Dirac \cite {D} that equations (\ref {h3})
can be written in the form
\bea \label{h35}
\frac {d} {dt}
\eta_n \approx  [\eta_n,H^T ]_D
\eea
where $f\approx g$ means that $f-g\in \Phi.$

Let $\{g\}\in A_D /\Phi$ be the coset represented by function $g.$
Then  equations (\ref{h35}) can be rewritten as
\bean \label{dr}
\frac {d} {dt}
\{\eta_n\}=\{ [\eta_n,H^T ]_D \}.
\eean
In $A_D/\Phi$ \bea
 \label {hof}
\{[g,f]_D\}=[\{g\},\{f\}]_D
\eea
and we get
\bea \label{h37}
\frac {d} {dt}
\{\eta_n\}=[\{\eta_n\},\{H\} ]_D.
\eea

In what follows we shall use first class
functions.
$R(\eta)$ is called a first class function \cite {D} if
\bean \label{} [R,\varphi_j
]|_{\varphi=0} =0 \eean or, equivalently,
 \bea \label{U} [
R,\varphi_j]=r_{jk}(\eta)\varphi_k .\eea
From equation (\ref {hh12}) it follows that  $H^T$ is first class.
It is known \cite {D} that first class functions form an algebra
with respect to the Poisson bracket. We shall denote this algebra by
$\Omega.$

\bigskip

{\bf 3.}
Let $\Upsilon$ be the space of the first
class functions which vanish on constraint surface
\bea \label{Did1}
\Upsilon =\{u\in \Omega\,|\,u=u_j(\eta){\varphi_j}\}.
\eea
Using (\ref {U}) and (\ref {Did1}) for $u\in \Upsilon ,g\in \Omega,$
   we get
\bean \label{Pid2} [u,g]|_{\varphi=0}=0.
\eean
From this and definition (\ref {Did1}) it follows
\bea \label{XU} [u,g] \in \Upsilon.
\eea
Hence $\Upsilon$ is an ideal of $\Omega$ \cite {L} and
$\Omega /\Upsilon$ is an algebra .

At this stage we have two quotient algebras $A_D /\Phi$ and
$\Omega /\Upsilon.$ Our aim is to show that they are isomorphic.

Let us define the linear function $T:
\Omega /\Upsilon \to A_D /\Phi$
\bea  \label {T}
T(\{g\}^\bullet)=\{g\}.
\eea
Here
$\{g\}^\bullet\in
\Omega /\Upsilon$ is the coset represented by $g.$
Since  $\{g\}^\bullet\subset \{g\},$
$T$ does not depend on the choice
of $g\in \{g\}^\bullet.$

To each function $g\in A$ one can put into correspondence the
first class function
\bean \label{fc} \tilde g=
g -[g,\varphi_{j}]c_{jk}\varphi_{k} .
\eean

Let $g'\in \{g\}$ be a first class function.
It can be written in the form
\bea  \label{ge}
g'=g-[g,\varphi_{j}]c_{jk}\varphi_{k}+v_j\varphi_{j}
\eea
where $v_j=v_j(\eta)$ are some functions.
Substituting $g'$ into
 the equation
\bean
[  g',\varphi_{k}]|_{\varphi=0}=0,
\eean
we obtain
\bean
[v_j\varphi_{j},\varphi_{k}]|_{\varphi=0}=0.
\eean
From this and equation (\ref {ge}) it follows that each first class
function $g'$ from $\{g\}$ is in
$\{g-[g,\varphi_{j}]c_{jk}\varphi_{k}\}^\bullet.$
This allows us to
define the inverse function $T^{-1}:A_D /\Phi \to \Omega /\Upsilon $
\bea \label {T-1} T^{-1}(\{g\})=\{g -
[g,\varphi_{j}]c_{jk}\varphi_{k}\}^\bullet .\eea
Thus we have shown that
$T$ defines the one-to-one correspondence between elements of $\Omega
/\Upsilon$ and $A_D /\Phi.$

In order to show that $T$ is a
homomorphism let us compute
$T([\{g\}^\bullet,\{f\}^\bullet]).$ Using definitions of $\Omega
/\Upsilon$ and $T$ we have \bea  \label {subalg}
T([\{g\}^\bullet,\{f\}^\bullet])=T(\{[g,f]\}^\bullet)=\{[g,f]\}.
\eea

Since $g$ and $f$ are first class,
\bean  \label {balg}
[g,\varphi_{j}] =g_{jj'}(\eta)\varphi_{j'},\qquad  [f,\varphi_{k}]=
f_{kk'}(\eta)\varphi_{k'}
\eean
and
\bean  \label {}
[g,f]_D=[g,f]+g_{jj'}\varphi_{j'}c_{jk}f_{kk'}\varphi_{k'}.
\eean
From this it follows that the right hand side of (\ref {subalg}) can be
written as \bea  \label {DD} \{[g,f]\}=\{[g,f]_D\}.  \eea
Using equation (\ref {hof}) and
definition (\ref {T}) we get
\bea
\label {mom}
\{[g,f]_D\}=[\{g\},\{f\}]_D= [T(\{g\}^\bullet),T(\{f\}^\bullet)]_D.
\eea

From equations (\ref{subalg})-(\ref{mom}) it follows that
\bea
\label {mom2}
T([\{g\}^\bullet,\{f\}^\bullet])=
[T(\{g\}^\bullet),T(\{f\}^\bullet)]_D
\eea
and hence $T$ is a homomorphism.
Since $T$ is an invertible function, algebras
$A_D /\Phi$ and  $\Omega /\Upsilon$ are isomorphic.

Elements of $\Upsilon$ act
in $\Omega$
as constraints.
Due to (\ref {XU}), for $u,v\in
\Upsilon $
\bean  \label {first}
[u,v] \in \Upsilon .
\eean
Hence constraints $\Upsilon$
are first class.
Thus we have shown that the Dirac bracket algebra $
A_D /\Phi$ is equivalent to the algebra of first class
finctions $\Omega$  subject to first class
constraints
$\Upsilon.$

Equation (\ref {XU})  tells us that
all the functions  of $\Omega$ are first class with respect to
constraints $\Upsilon.$
From this it follows that the system under consideration
is invariant with respect to the gauge transformations
\bean \label{} \delta_{u}g
=[u,g] \eean where $u\in \Upsilon,\,g\in \Omega.$

According to (\ref {T-1}) and (\ref {mom2}) the image of the Hamilton
equations (\ref {h37}) in $\Omega /\Upsilon$ is \bean \label{h77} \frac
{d} {dt} \{{\tilde \eta}_n\}^\bullet= [\{{\tilde
\eta}_n\}^\bullet,\{H^T\}^\bullet].
\eean Here ${\tilde \eta}_n =\eta_n -
[\eta_n,\varphi_{j}]c_{jk}\varphi_{k} .$

Let $\tilde A=A[[\lambda]]]$
be the space of formal power series in a variable $\lambda$
with coefficients in $A$. Let $*$ be the star-product \cite {St}
\bea
g(\lambda)*f(\lambda)=gf+\sum_{k=1}^\infty \lambda^kc_k,
\eea
where $g(\lambda)=g+\sum_{k=1}^\infty \lambda^kg_k, f(\lambda)=f +
\sum_{k=1}^\infty \lambda^kf_k$  and  $g,f,c_k,g_k,f_k\in A.$

Using the $*-$product one defines the $*-$commutator
algebra  $\tilde A_c$
\bea \label {ca}
[g,f]_c=\frac 1 {2\lambda} \left( g(\lambda)*f(\lambda)-f(\lambda)*g(\lambda)
\right)=[g,f] +
\sum_{k=1}^\infty \lambda^kb_k, \eea where $b_k\in A.$
From the definition it follows that $\lambda\tilde A_c$ is an ideal of
$\tilde A_c$ and $\tilde A_c/\lambda \tilde A_c$ is the algebra which is
isomorphic to the Poisson bracket algebra $A_P.$ Let $L: A_P  \to
\tilde A^c/ \lambda\tilde A^c$ be the corresponding map.  Then the
isomorphic image of $\Omega /\Upsilon$ (and $A_D/\Phi$) in $\tilde
A^c/\lambda \tilde A^c$ is $L(\Omega) /L(\Upsilon).$ Thus we have
obtained the realization of the Dirac bracket algebra $A_D/\Phi$ as a
subalgebra of the $*-$commutator algebra $\tilde A^c/\lambda \tilde
A^c.$

\bigskip

I wish to thank D.Sternheimer for usefull correspondence.


\bigskip
\begin{thebibliography}{99}
\bibitem {D}P.A.M.Dirac,
{\it Lectures on Quantum Mechanics} (Yeshiva University, New York,1964).
\bibitem {BFF}I.A.Batalin,E.S.Fradkin and T.Fradkina,{\it
Nucl.Phys.} {\bf 332} (1990) 723.
\bibitem {S}I.J.Sniatycki,{\it Ann.Inst.H.Poincare
} {\bf 20} (1974) 365.
\bibitem {L}A.Lichnerowicz,{\it C.R.Acad.Sci.
Paris} {\bf A 280} (1974) 523.
\bibitem {FLS} M.Flato,A.Lichnerowicz,and D.Sternheimer {\it
J.Math.Phys.} {\bf 17} (1976) 1754.
\bibitem {F} A.T.Fomenko,
{\it Symplectic geometry} (Moscow University,Moscow
,1988) (in Russian).
\bibitem{B} A.V.Bratchikov,{\it Mod.Phys.Lett.}
{\bf A16} (2001) 1139.
\bibitem{St} D.  Sternheimer,{\it \ ``Deformation
Quantization:  Twenty
Years After'',} math.QA/9809056.
 \bibitem {GT} D.M.Gitman and
I.V.Tyutin, {\em Canonical Quantization of Fields
with Constraints}
(Nauka, Moscow, 1986)(in Russian).  \end{thebibliography}
\end{document}